\documentclass[12pt,a4paper,epsf]{article}
\usepackage{graphics}
\usepackage{amssymb,amsmath}
\usepackage[dvips]{lscape,graphicx}
\usepackage{cite}
\usepackage{longtable}
\usepackage{slashed}

\newcommand{\ct}{\cite}
\newcommand{\lb}{\label}

\newcommand{\bc}{\begin{center}}
\newcommand{\ec}{\end{center}}
\newcommand{\bd}{\begin{displaymath}}
\newcommand{\ed}{\end{displaymath}}
\newcommand{\be}{\begin{equation}}
\newcommand{\ee}{\end{equation}}
\newcommand{\ba}{\begin{array}}
\newcommand{\ea}{\end{array}}
\newcommand{\bea}{\begin{eqnarray}}
\newcommand{\eea}{\end{eqnarray}}
\newcommand{\bt}{\begin{tabular}}
\newcommand{\et}{\end{tabular}}

\newcommand{\ov}{\overline}
\newcommand{\bp}{\begin{picture}}
\newcommand{\ep}{\end{picture}}
\newcommand{\bfi}{\begin{figure}}
\newcommand{\efi}{\end{figure}}

\def\fun#1#2{\lower3.6pt\vbox{\baselineskip0pt\lineskip.9pt
\ialign{$\mathsurround=0pt#1\hfil##\hfil$\crcr#2\crcr\sim\crcr}}}

\begin{document}



\vspace{1cm}

\title{\LARGE {\bf Diphoton decay of the Higgs boson and new bound
states of top and anti-top quarks}}
\author{\large \bf C.D.~Froggatt$\large
\bf{}^{1}$\large \bf \footnote{c.froggatt@physics.gla.ac.uk}\,,
C.R.~Das$\large \bf{}^{2}$\footnote{crdas@cftp.ist.utl.pt
}\,,
L.V.~Laperashvili$ \large \bf{}^{3}$\large \bf \footnote{laper@itep.ru}\, and
H.B.~Nielsen$\large \bf{}^{4}$\large \bf \footnote{hbech@nbi.dk, hbechnbi@gmail.com}\\[5mm]
 \itshape{ $
\large \bf{}^{1}$ Department of Physics and Astronomy,
 Glasgow University,
 Scotland}\\
\itshape{$\large \bf{}^{2}$ Theory Division, Physical Research Laboratory, 
Ahmedabad,
India}\\
\itshape{\ $ \large \bf{}^{3}$ Institute of Theoretical and
Experimental Physics, Moscow, Russia}\\
\itshape{$ \large \bf{}^{4}$ The Niels Bohr Institute, Copenhagen,
Denmark }}

\date{}

\maketitle

\thispagestyle{empty}

\begin{abstract}

We consider the constraints, provided by the LHC results on Higgs
boson decay into 2 photons and its production via gluon fusion,
on the previously proposed Standard Model (SM) strongly bound
state $S$ of 6 top quarks and 6 anti-top quarks. A correlation
is predicted between the ratios $\kappa_{\gamma}$ and $\kappa_g$ of
the Higgs diphoton decay and gluon production amplitudes respectively
to their SM values. We estimate the contribution to these amplitudes
from one loop diagrams involving the 12 quark bound state $S$ and
related excited states using an atomic physics based model. We find
two regions of parameter space consistent with the ATLAS and CMS data
on ($\kappa_{\gamma}$, $\kappa_g$) at the 3 sigma level: a region
close to the SM values ($\kappa_{\gamma}=1$, $\kappa_g =1$)
with the mass of the bound state $m_S > 400$ GeV and a region
with ($\kappa_{\gamma} \sim 3/2$, $\kappa_g \sim -3/4$) corresponding
to a bound state mass of $m_S \sim 220$ GeV.

\end{abstract}

\thispagestyle{empty}

\clearpage\newpage

\section{Introduction}

\label{introduction}

Previously we have speculated \ct{1,2a,3,4,5,6,8,9,10,11,12,13}
that 6 top + 6 anti-top quarks should be so strongly bound that the
bound states, having masses small compared to the total mass of 12 top quarks,
would effectively function as elementary particles to first approximation.
In this case they can be added into loop calculations as ``new'' particles
in the theory and seen, if produced at LHC, as resonances.
Now the Higgs boson decay into two photons and its production via
the gluon fusion mechanism are loop induced processes in the
Standard Model\footnote{Here of course we refer to the usual
Standard Model one loop diagrams involving just the fundamental
SM particles. In this paper we are claiming that the true
Standard Model calculation should include the loop contributions
from our new bound states.} (SM) and therefore relatively suppressed.
So both these processes should be sensitive to loop contributions
from our proposed new bound states.

These new bound states are supposedly held together by the exchange of
Higgs bosons and gluons between all the constituents.
This is because the Higgs field leads to attraction between
$ tt,\,t\bar t$ and $\bar t\bar t$  quarks and the Yukawa coupling
constant $g_t$ in the Lagrangian, describing the interaction of
top-quarks with the Higgs boson:
\be
 L = \frac 12 D_{\mu}\Phi_H D^{\mu}\Phi_H + \frac{g_t}{\sqrt
          2}\ov{\psi}_{tL}\psi_{tR}\Phi_H  + h.c. \lb{1} \ee
is large enough: ${g_t\sim 1}$.
If these bound states are indeed very light compared to the
sum of the constituent masses, we expect effects arising from them
which are not included in the usual type of essentially
perturbative calculations using only the fundamental particles
of the SM as elementary particles.

In the present article we consider these effects in the amplitudes
for Higgs diphoton decay and for Higgs production via gluon fusion.
In fact we estimate the scale factors ($\kappa_{\gamma}$ and $\kappa_g$)
by which the usual SM amplitudes for these processes have to be
scaled when the effects of our proposed strongly bound states are
included.

The ATLAS and CMS collaborations have extracted
values for these scale factors from their LHC data:
\be \kappa_{\gamma} = 0.97 \quad (2\sigma\,\,{\rm interval}\,\,
0.59\,\, {\rm to}\,\, 1.30), \qquad
%
%
\kappa_g = 0.83 \quad (2\sigma \,\, {\rm interval}\,\,
0.63\,\, {\rm to}\,\, 1.03) \lb{K1} \ee
%
for CMS \ct{19}, and
\be \kappa_{\gamma} = 1.20 \pm 0.15, \qquad
%
%
\kappa_g = 1.04 \pm 0.14 \lb{K2} \ee
%
for ATLAS \ct{20,20a}.
These experimental values are consistent with the usual SM values
$\kappa_{\gamma} = 1$, $\kappa_g =1$, but are also consistent
with significant corrections to the usual SM values.
We confront our estimates of the corrections from our new bound
states with these data and discuss the implications for our model.

In estimating these corrections from our strongly bound states,
we have to make a number of severe approximations:
\begin{enumerate}
\item
We use non-relativistic calculations to evaluate the interaction vertices
involving the bound states, in order to avoid Bethe Salpeter equations.
This also means that we ignore the finite speed of Higgs exchange
and use an instantaneous Higgs exchange potential. Consequently the
spin of the quarks effectively becomes an internal decoupled degree of
freedom for our calculation.

\item
We ignore the Higgs mass and use a Coulomb potential in first approximation;
in an earlier article \ct{10} we argued that, inside our strongly bound states,
the Higgs field would be strongly reduced compared to its usual vacuum
expectation value and that, consequently, the effective Higgs mass inside the
bound state would be considerably smaller than the physical Higgs mass.
It is this zero Higgs mass assumption combined with non-relativistic physics
that allows us to use Bohr model estimates for the interaction vertices
involving the bound states. In appendix A we give a procedure for obtaining
relativistic expressions for the corresponding interaction terms in the
Lagrangian density from these non-relativistic coupling results.

\item
When we look at just one out of the 12 quarks or anti-quarks in the potential
created by the other 11, typically one half of these 11 are actually on the
outside of the quark considered. We correct for this ``screening" effect by
just taking the central charge in our Bohr model calculation to be as if there
were 11/2 quarks or anti-quarks sitting at the centre of the bound state.

\item
We ignore the forces between the quarks and anti-quarks due to the exchange of
gluons and eaten Higgs (longitudinal $W^{\pm}$ and $Z^0$) particles in our Bohr
model calculation. These forces and several other corrections were included in
the calculation in Ref.~\ct{10} of the strong binding of the ground state $S$
of 6 top and 6 anti-top quarks. In this reference the radius of the bound state
$S$ was estimated to be $\sim 1/m_t$ ($m_t$ is the top quark mass) and thus of a
similar magnitude to the Compton wavelength of $S$.

\item
Most importantly we assume that the bound states can, in first approximation,
be considered as {\em point particles} and that it is sufficient to then
introduce crude form factor corrections. However, in our Bohr model calculation,
the radius r of the bound state $S$ (formally estimated in appendix B as
$r \sim 5/m_t$) is large compared to its Compton wavelength as would be given
by the masses for $S$ considered in this paper, e.g.~the estimate
$m_S \sim 260$ GeV given in Ref.~\ct{52}. Therefore $S$ is not approximately
point-like in our Bohr model. Consequently the form factor
corrections become embarrassingly large. But we show in appendices B and C that
our estimate of the loop amplitude $A_S$ for the bound state contribution to
Higgs diphoton decay is rather insensitive to the radius $r$; naively
the bound state polarizability $\alpha_{pol.} \sim r^3$ while the effect
of the form factor varies as $1/r^4$, so that the two terms largely
compensate each other. However, in the Bohr model, $\alpha_{pol.}$ also has a
dependence on the central charge $Z$ on the ``nucleus". If the radius is
imagined to vary as a result of varying $Z$, keeping the top quark mass $m_t$
constant, then $\alpha_{pol.} \sim m_t r^4$ and completely compensates the $1/r^4$
dependence from the form factor. We can therefore hope that the inclusion of the
corrections in Ref.~\ct{10}, leading to $r \sim 1/m_t$, would leave our estimate
of the loop amplitude $A_S$ essentially unchanged.
The reduction of $A_S$ by our form factor correction
would then be a much smaller effect and treating $S$ as a point particle in
first approximation would be more realistic.

\end{enumerate}

In section 2 we present our motivation, based on the so-called
multiple point principle, for proposing the existence of these new bound
states in the Standard Model. We then briefly review our calculation \ct{10} of
the strong binding of the scalar 1s ground state $S$ of 6 top and 6 anti-top
quarks. We also consider the first excited states of this scalar bound state
and the charged spin-1/2 bound state $F$ of 6 top and 5 anti-top quarks.
The possible production of these new bound states at the LHC is also discussed.

In sections 3 and 4 we turn to the main content of the present article - the
evaluation of the contribution of our $S$ and $F$ bound states\footnote{These
bound states have also been called \ct{8,11,12,13} $T_s$-balls and $T_f$-balls
respectively.} to diphoton
Higgs decay and Higgs production via gluon fusion respectively. The scalar
bound state $S$ is a self-conjugate neutral object, which does not
couple directly to photons or gluons. However it can couple to them by
being polarized, meaning that it is virtually excited by, say, a photon
into a higher energy state and then de-excited by a second photon.
We take the virtual intermediate state $S_1^*$ to be a spin-1 state,
in which one of the top quarks in the ground state $S$ has been
excited to a 2p orbit. The general features of the decay and production
amplitudes arising from the loop diagrams including the usual SM particles
and the above new bound states are presented. The predicted correlation between
$\kappa_{\gamma}$ and $\kappa_g$ is discussed and confronted with
the experimental data in section 5.

Section 6 contains a discussion of the approximation of treating our new bound
states as effectively elementary particles and the needed corrections
to it. We estimate the form factor corrections to be included in the loop
diagrams involving these bound states in terms of their radii.
In section 7 we evaluate the contribution of the scalar $S$ and spin-1/2 $F$
loop amplitudes to the scale factors $\kappa_{\gamma}$ and $\kappa_g$.
We find that the fermionic $F$ loop is strongly damped and can be neglected.

The results of our model calculations are compared with the LHC data
in section 8 and our conclusions are presented in section 9.

We relegate some details of our calculation to a series of appendices.
In appendix A we present our prescription for the normalization of
vertices involving the scalar bound state $S$. The polarizability
of the $S$-particle is estimated in appendix B, by analogy with
that of the hydrogen atom. Finally, in appendices C and D, we give the
details of the calculation of the scalar $S$ and fermionic $F$ loop
amplitudes respectively.

\section{The new bound states}

The crux of the present paper is that {\it in the pure Standard Model}
there may exist bound states of six top quarks and six anti-top quarks,
which are expected to show up at the LHC. Indeed we imagine that the SM
could turn out to be valid up to essentially
the Planck scale, apart from the existence of some right-handed neutrinos
at a see-saw scale of around $10^{12}$ GeV. This scenario of course suffers
from the hierarchy problem of why the ratio of the Planck scale to
the electro-weak scale is so huge. We have suggested a mechanism for fine-tuning
the coupling constants in the SM \ct{1,2a,3,4,5,6}
so that the ratio of these scales should
be of the order of $10^{-17}$, as observed, based on the so-called Multiple
Point Principle \ct{15x,16x,17x,19x,18x,22x,29x,32x}.

According to the Multiple Point Principle (MPP), there should exist several vacua
having the same energy density. This principle of degenerate vacua was applied
some time ago \ct{33x} to the SM, by assuming the existence of a second vacuum with
a Higgs field vacuum expectation value (vev) close to the Planck scale and
degenerate with the usual SM vacuum having a vev of $v= 246$ GeV. Consequently the
top quark and Higgs boson couplings were fine-tuned to lie at a point on the SM vacuum
stability curve\ct{40,41,42,43,44,45} corresponding to the Planck scale. This
led to the following MPP prediction \ct{33x} for the masses of the top
quark and the Higgs boson:
$$m_t = 173\pm 5\,\,{\rm GeV},\quad m_H = 135 \pm 9\,\,{\rm GeV}.$$
There now exist next-to-next-to leading order 2-loop calculations \ct{35x,36x,34x,37x}
of the SM vacuum stability curve. Using a top quark pole mass of
$m_t = 173.1 \pm 0.7$ GeV as input leads \ct{37x} to an updated MPP prediction
for the Higgs mass of $m_H = 129.4 \pm 1.8$ GeV. This is remarkably close
to the observed Higgs mass \ct{19,20,20a} of 125-126 GeV and we note that
it is rather sensitive to the top quark pole mass; a change of
$\Delta m_t = \pm 1$ GeV gives a change in the MPP predicted Higgs mass of
$\Delta m _H = \pm 2$ GeV.

The intriguing result that, assuming the validity of the SM to very high scales,
the measured value of $M_H$ lies on or is very close to the vacuum stability
curve could of course be a pure coincidence. However we take the attitude that
it is not accidental and requires an explanation. This implies two points:
\begin{enumerate}
\item The SM should not be modified so much, by new physics in the energy
range between the electro-weak scale and the very high energy scale of the
second vacuum, that the renormalisation group running of the Higgs quartic
coupling $\lambda(\mu)$ is significantly altered.

\item There must for some reason exist in Nature a physical
principle forcing one vacuum to be so closely degenerate with
another one that it is barely stable or just meta-stable.
One such principle is of course the above-mentioned multiple
point principle.
\end{enumerate}

The multiple point principle (MPP) of several vacua having closely
the same energy density (in fact approximately zero energy density)
is really a mechanism for fine-tuning couplings (just by
formulating a rule for this fine-tuning, instead of seeking to
avoid it). In order to fine-tune the SM couplings so as to generate the large
ratio of the Planck scale to the electro-weak scale using MPP, it is necessary
\ct{1,2a,3,4,5,6} for them to produce a third essentially zero energy density vacuum
in the SM. In our speculative picture this new vacuum is formed at the
electro-weak scale by a Bose condensation of a strongly bound state of
6 top and 6 anti-top quarks. We note that 1 cm size balls of this new vacuum,
packed with white dwarf like normal matter highly compressed by the skin separating them
from the normal vacuum, could provide a viable model of dark matter purely
within the SM \ct{50,51,52}.

In Ref.~\ct{1} it was first proposed that\\
$\bullet$  there exists a
$1s$-bound state of $6t+6\bar t$ quarks, which is a scalar and
color singlet called $S$ in this article;\\
$\bullet$ the forces responsible for the forming of these
bound states originate from the virtual exchanges of the Higgses
(including the eaten components and also the gluon) between top quarks;\\
$\bullet$ these forces are so strong that they almost
compensate the mass of the
12 top quarks which are contained in these bound states;\\
$\bullet$ there also exists a new bound state of $ 6t + 5\bar
t$ quarks, which is a fermion similar to the quark of a 4th generation
having quantum numbers of the top quark, and called $F$ here.

For the Higgs field which like gravity is described by an even-order
tensor field (namely a scalar), there is the same rule as for
gravity that both particles and anti-particles attract both
particles and anti-particles. So as more top and/or
anti-top quarks are put together, the Higgs exchange binding energy is
essentially proportional to the number of pairs of constituents, rather
than to the number of constituents, and the stronger they bind.
However the top quarks are fermions and the Pauli principle restricts
the number that can be put into the $1s$ ground state.
Because the quark has three color states
and two spin states, meaning six internal states, there is in fact a
shell (in the nuclear physics sense) with six top quarks and
similarly one for six anti-top quarks. Like in nuclear physics,
where the closed shell nuclei are the strongest bound, we imagine
that when these shells are filled with 6 top and 6 anti-top quarks
we obtain the most strongly bound and also the lightest bound state $S$.
This bound state $S$ is expected to be stable w.r.t. the
emission of a top or anti-top quark, but it is very likely that it will be
able to decay by splitting into jets of lighter particles,
although probably with a decay rate which is surprisingly low for
its mass. Also excited $6t+6\bar t$ states, in which one of the quarks
is excited to a 2s or 2p level (in the atomic physics terminology),
should exist, forming a heavier scalar $S^*$ and vector particle
$S^*_1$ respectively.

Since, in our MPP scenario, there should exist a vacuum state
containing a condensate of the bound states, an effective field theory
for the bound state $S$ should have a tachyonic mass term with a
non-zero vacuum expectation value for the bound state field
$<\phi_S> \neq 0$. Moreover this vacuum should be degenerate with the
normal SM vacuum and thus be only barely formed. Consequently, although
formally tachyonic, the bound state mass term should be close to being
a normal positive mass. Thus we expect our bound state $S$ to be appreciably
lighter than its natural scale of 12 times the top quark mass, which is
about 2 TeV. In a recent paper \ct{52} the mass of the bound state $S$
was estimated to be $m_S \approx 260$ GeV, but really all we can say
without much more detailed calculations is that $m_S$ should be much smaller
than 2 TeV. There is a lower limit on the mass of $m_S > m_H/2 =63$ GeV,
since otherwise the Higgs particle would decay dominantly into pairs of
the bound state $S$.

The existence of a degenerate vacuum with a bound state condensate
of course implies a fine-tuning of the strength of the top quark coupling
to the Higgs field. Two of us have made \ct{10} a detailed analysis of this
fine-tuning condition and obtained an estimated value for the top quark
running Yukawa coupling constant of $g_t = 1.00 \pm 0.14$, which within errors
is consistent with the experimental value of 0.935. However this result is
controversial, as even the very binding of the bound state has been
disputed \ct{15,16,17,18}. So more accurate calculations of the binding
of $6t+6\bar t$ states, if possible on the lattice say, are highly
called for; it should be emphasized that, although complicated, these are
calculations purely within the SM where all the parameters are known.

Having taken the bound state $S$ of $6t+6\bar t$ quarks to be so strongly
bound as to be much lighter than the collective mass of its constituents,
we would also expect that making small
modifications of this bound state, such as removing one anti-quark
(or one quark) would still lead to a remarkably light bound state
or resonance. Actually it is not difficult to make a crude estimate
of a curve for how the mass of the bound state of a number of top quarks and
anti-top quarks, up to the closing of the first shell, will vary with
the number of constituents \ct{9,10}. Assuming that the full shell bound
state $S$ has a very small mass relative to the scale of 12 top quark
masses, this interpolating curve leads to a mass for a bound state of
11 top and anti-top quarks (called $F$ earlier in this section)
of around 760 GeV.  One should have in mind that the binding of the
bound state $S$ is so strong in our speculation
that the last quark out of the 12 binds with a binding
energy larger than its mass energy. So the spin-0 bound state $S$
of 12 constituents is supposed in our picture to be {\em
lighter} than the spin-1/2 bound state $F$ with only 11 constituents.

\bfi\centering\includegraphics[height=90mm,keepaspectratio=true,angle=0]
{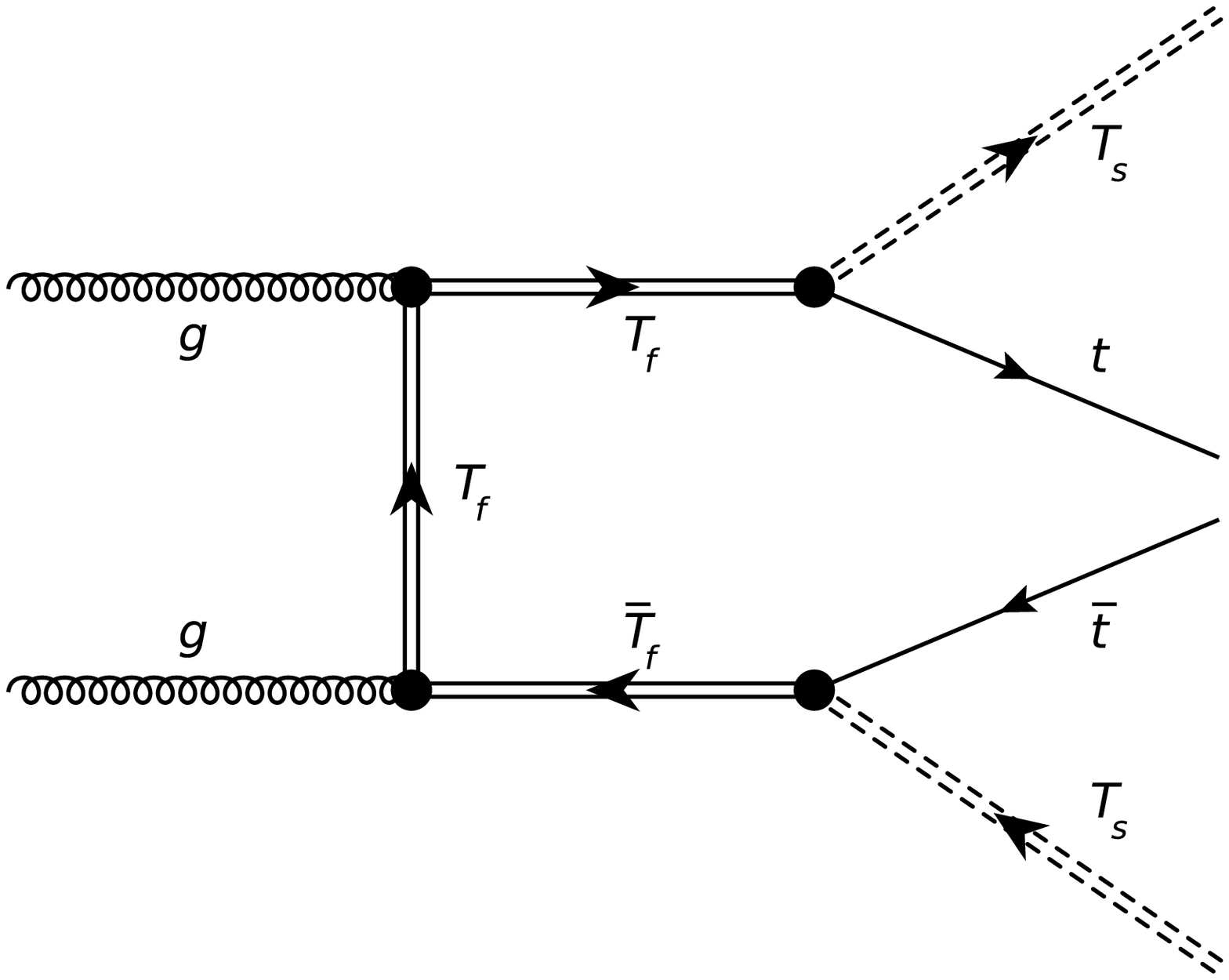}\caption{Feynman diagram for producing a pair
of  charged spin-1/2 bound states $F^{\pm}$ with only ''11''\, constituent
$t$ and $\bar t$ quarks in them. Here $T_s$ and $T_f$ denote the scalar
bound state $S$ and the spin-1/2 bound state $F$ respectively.} \lb{F1}\efi

The bound state $S$ is colorless and therefore couples only weakly to gluons.
It can couple by being polarized, meaning that it is brought into an excited state
virtually by a gluon and then, by coupling to a second gluon, gets
de-excited to the original bound state $S$ again. In principle,
this polarization coupling can give rise to the production of two
bound states $S$ by the collision of gluons and should be looked for.
If one assumes that the bound state
$S$ decays into a couple of hadronic jets, which is quite
likely, then one should look for four jet events at the LHC,
in which two pairs of the jets would each form $S$ resonances with the
same mass $m_S$. If this mass is small, it might be seen as
just a two-jet event.

An alternative production mechanism of the bound state $S$ is via the
production and decay of the heavier spin-1/2 bound state $F$, which
couples directly to gluons. We therefore imagine that the major
production mechanism is by pair production of $F$ resonances, each of
which then successively decay to the spin-0 bound state $S$ and a top
or anti-top quark. This process is illustrated by the diagram in Fig.~1.
This diagram is very analogous to the diagram for pair producing a fourth
generation $b'$ quark. In fact such a fourth generation $b' \bar b' $ quark
pair can be produced by gluon collision followed by the decays
$b' \to W^- t$ and $\bar b' \to W^+ \bar t$ respectively. So the
production of our bound states $S$ would look very similar to the
production of $b' + \overline{b'}$ provided the W
could be confused with our bound state, as could easily happen in
as far our bound state presumably decays into two hadronic jets which
can also easily happen for $W$'s. Of course, however, the mass of
our bound state $S$ is not expected to be just that of a
$W$-boson. So events containing the new bound states should be
distinguishable from $b' + \overline{b'}$ by the mass of
the $W$-analogous particle (namely the spin-0 bound state $S$)
being sharp but presumably different from that of $W$.
The top quarks in the production mechanism of Fig.~1 decay into
$W$'s and $b$ quarks. If the $W$'s then decay hadronically and the
$b$'s also decay hadronically, showing up simply as single jets
at these high energies, the whole event will show up as a rather
clear 10 jet event.

If we take the analogy with the pair production of a fourth generation
quark seriously, we expect that apart from form factor effects the rate of
pair production of the spin-1/2 bound states $F + \bar F$ should be very
similar to that of the production rate for a pair of fourth generation
$t' + \bar t'$ quarks\footnote{We note that a chiral doublet of fourth
generation quarks would give a contribution to Higgs diphoton decay
corresponding to scale factors of $\kappa_{\gamma} = 0.65$ and
$\kappa_g = 2.25$, which are inconsistent with the data - see Fig. 3.
However a fourth generation of vector-like quarks would be allowed, as
they would not couple to the Higgs particle.}.
These form factor effects are somewhat uncertain,
although in principle they could even give an enhancement of the
production rate, since the gluons producing a $F + \bar F$ pair
would be a bit time-like. However, if the bound state $F$ is not sufficiently
strongly bound and light, it may not be a good approximation to treat it
as a fundamental particle in analogy to a fourth generation. Should this
approximation fail, the cross section for the production of an $F + \bar F$
pair could be much lower than estimated by analogy to a fourth generation
quark pair production rate.

\section{The decay width of $H\to\gamma\gamma$}

The diphoton decay of the Higgs boson $H$ is a loop induced process. In the usual
SM the process is dominated by the $W^{\pm}$ and top quark loops, with the
$W^{\pm}$ loop amplitude approximately 4$\frac{1}{2}$ times bigger than the $t$ quark loop
amplitude and of opposite sign. The corresponding Feynman diagrams are shown
in Fig.~2a.

\bfi\centering\includegraphics[height=50mm,keepaspectratio=true,angle=0]
{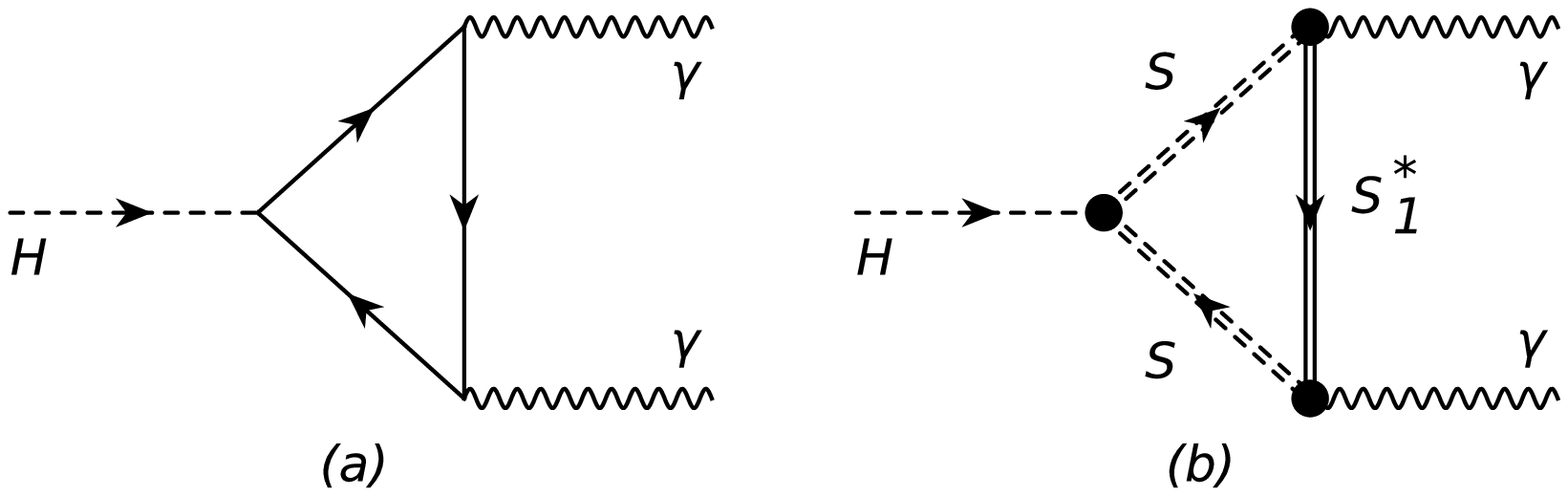}\caption{The one-loop triangle diagrams
describing the contributions of different particles to the
diphoton decay of the Higgs boson: (a) describes the contributions
of $W,\,t$ and the spin-1/2 bound state $F$ and (b) describes the
contribution of the scalar bound state $S$ via the dipole
interaction $SS^*_1\gamma$.}
\lb{F2}\efi

The analytic expression for the usual SM diphoton partial decay width
of the Higgs particle reads
\ct{28,29,30,31} as follows:
\be
 \Gamma_{SM}(H\to\gamma\gamma) =
 \frac{G_F\alpha^2m_H^3}{8\sqrt 2\pi^3} |A_W(\tau_W) +
 N_cQ^2_tA_t(\tau_t)|^2,  \lb{20}    \ee
where $G_F$ is the Fermi constant, $N_c = 3$ is the number of
colours, $Q_t = +2/3$ is the top quark electric charge in units of
$|e|$, $\alpha=e^2/4\pi$, and
\be
\tau_i = 4m^2_i /m^2_H.
\lb{20a} \ee
In this section and below we use the notation $m_i$ for the masses of
the particles $i=H,\,W,\,t,\,F$ and $S$.

Below the $WW$-threshold, the loop functions for spin-1 (W boson)
and spin-1/2 (top quark) are defined as follows: $A_W=A_1(\tau_W)$
and $A_t=A_{1/2}(\tau_t)$, where (see Refs.~\ct{28,29,30,31})
\be A_1(x) = - \frac{x^2}{4}[2x^{-2} + 3x^{-1} + 3(2x^{- 1} -
1)f(x^{-1} )],
\lb{21} \ee       
\be A_{1/2}(x) = \frac{x^2}{2}[x^{-1} + (x^{-1} - 1)f(x^{-1})],
                                              \lb{22}      \ee
\be f(x^{-1}) = ({\rm arc\,\, sin} \sqrt {x^{-1}})^2.  \lb{22a}
\ee
Here we use the normalization of Ref.~\ct{28}. In the limit when
the particle running in the loop has a mass much heavier than the
Higgs, we have:
\be A_1 \to - \frac 74, \quad  N_cQ^2_tA_{1/2}\to \frac 13
N_cQ^2_t.      \lb{23}  \ee                                       %
The W boson provides the dominant contribution to
$\Gamma_{SM}(H\to\gamma\gamma)$, while the top quark
contribution is well-approximated by the asymptotic value of
$\frac 13 N_cQ^2_t = 4/9$. We take a Higgs mass of 126 GeV,
for which the W and top contributions are:
\be m_H \approx 126\,\, {\rm GeV}, \quad A_W \approx -2.08, \quad
N_cQ^2_tA_t\approx 0.46.   \lb{24} \ee

We now wish to introduce the effects on the diphoton decay width
from adding loop diagrams involving the new bound states,
which would interfere with the SM contributions.
Assuming there is only the SM Higgs boson, we start by re-writing
the diphoton decay width in terms of couplings $G_{HWW},\,G_{Ht\bar
t},\,G_{HF\bar F},\,G_{HSS}$, which are the Higgs couplings to the
particles in the loop diagrams of Fig.~2:
$$
 \Gamma(H\to\gamma\gamma) =
 \frac{m_H^3}{64\pi^3} |\alpha \frac {G_{HWW}}{m_W}A_W(\tau_W) +
 2\alpha \frac{G_{Ht\bar t}}{m_t} N_cQ^2_t A_t(\tau_t)$$ \be
  + 2\alpha \frac{G_{HF\bar F}}{m_F} N_cQ^2_t A_F
   + \alpha \frac{G_{HSS}}{m_tm_S}A_S|^2.
    \lb{25} \ee
In the above equation, the notations $F$ and $S$ refer to the spin-1/2
and spin-0 new bound states, respectively. If we could treat the bound
state $F$ as an elementary particle, the amplitude $A_F$ would be
given by
%
$A_F = A_{1/2}(\tau_F) \approx 1/3.$
The amplitude $A_S$ corresponds to the loop diagram of Fig.~2b
and arises from the polarizability of the self-conjugate scalar bound
state $S$, due to its dipole interaction $SS^*_1\gamma$ with the spin-1
excited state $S_1^*$. If we could treat the bound states $S$ and $S_1^*$
as elementary particles, the amplitude $A_S$ would diverge.
Our evaluation of the polarizability and the
amplitude $A_S$ is discussed in Section 7 with more details given
in the appendices.

The electric charge of the fermion $F$ in units of
$|e|$ is $Q_F=Q_t$ and $N_c=3$ is the number of colors. So, in
Eqs.~(\ref{20}) and (\ref{25}) we have:
\be N_cQ^2_t = \frac 43. \lb{26} \ee
Using the well-known SM relations $$v^2 = \frac{1}{\sqrt{2}
G_F}, \quad m_t=\frac{g_t}{\sqrt
2}v,$$
%
the $W$ boson and top quark couplings to the
Higgs are given by
\be \frac{G_{HWW}}{m_W} = 2\frac{G_{Ht\bar t}}{m_t} = \frac{2}{v}.
\lb{32} \ee

We use an ``impulse approximation" to calculate the couplings of the
Higgs field $H$ to the 11 quark bound state $F$ and the 12 quark
bound state $S$, in which we add the
interactions of $H$ with each of the $t$ and $\overline{t}$ quarks
individually as if they were free particles. The coupling to $F$
is then given by
\be G_{HF\bar F} = 11\frac{g_t}{\sqrt 2} =
11\frac{m_t}{v}. \lb{33}   \ee
As described in appendix A, we now obtain the coupling constant
between the three scalar particles $SSH$ by
comparing with a non-relativistic interaction of the $6t+6\bar t$
bound state $S$ with the Higgs field $H$:
%
\be G_{HSS} = 12\frac{g_t}{\sqrt{2}} \cdot 2m_S = 24 \frac{m_tm_S}{v}.
\lb{34} \ee

Finally we have:
\be
 \Gamma(H\to\gamma\gamma) =
 \frac{\alpha^2m_H^3}{16\pi^3v^2} |A_W(\tau_W) + \frac 43 A_t(\tau_t)
  + \frac{44}{3}\frac {m_t}{m_F}A_F + 12A_S|^2.
    \lb{35} \ee
which can be rewritten in the form
\be
 \Gamma(H\to\gamma\gamma) =
 \frac{\alpha^2m_H^3}{16\pi^3v^2} |A_W(\tau_W) + \frac 43 A_t(\tau_t)
  - K|^2.
    \lb{35a} \ee
Here, for convenience, we define
\be
K = -\left(\frac{44}{3}\frac {m_t}{m_F}A_F + 12A_S\right)
\lb{35b} \ee
to be the negative of the total contribution from the bound states $F$ and $S$..

Now the ratio of the diphoton decay width $\Gamma(H\to\gamma\gamma)$ in our model
and the usual SM decay width $\Gamma_{SM}(H\to\gamma\gamma)$ is just equal
to the square of the scale factor $\kappa_{\gamma}$ in our model.
So, forming the ratio of Eqs.~(\ref{35a}) and (\ref{20}), we obtain
\be
\frac{\Gamma(H\to\gamma\gamma)}{\Gamma_{SM}(H\to\gamma\gamma)} =
\kappa_{\gamma}^2 = \left(1 - \frac{K}{A_W + \frac 43 A_t}\right)^2
  \lb{20b}  \ee
Substituting the values of $A_W$ and $\frac{4}{3}A_t$ for a Higgs mass of
$m_H = 126$ GeV from Eq.~(\ref{24}) into Eq.~(\ref{20b}), we have
\be
\kappa_{\gamma}^2 = \left(1 + \frac{K}{2.08 - 0.46}\right)^2
  \lb{20c}  \ee
and thus finally we obtain the result
\be
\kappa_{\gamma} = 1 + \frac{K}{1.62}.
  \lb{20d}  \ee

\section{Gluon fusion production}

The production of the Higgs boson is dominantly via the so-called gluon
fusion mechanism. The corresponding Feynman diagram in the usual SM
is a triangle diagram having two incoming gluons from the colliding
(say, LHC) protons and one Higgs boson attached to the vertices, with a top
quark circling around the loop.
Now when our new bound states are introduced, the gluon fusion
production amplitude comes to depend on them, somewhat by accident, in a
similar way to the dependence on them of the Higgs boson decay amplitude into
two $\gamma$'s, which we considered in Section 3. The point is that the only
difference between the diagrams for the decay $H\to\gamma\gamma$ (see Fig.~2)
and the diagrams (time reversed) for the gluon fusion production $gg \to H$
is that the two external photons are replaced by two external gluons. Since
both gluons and photons are vector-particles, the contraction in
the geometrical index $\mu$ becomes completely analogous and only
the ``charges'', the color and charge dependent couplings, are
modified in going from the diphoton decay diagram to the gluon
fusion one. Now the constituents of our bound states which really couple to
the photons and gluons respectively in the loop diagrams are
just top and anti-top quarks. So the ratio of their couplings to the photons
and the analogous gluons is a quite fixed ratio equal to that of the top
quark in the usual SM loop contribution. The W-boson of course does not contribute
to the gluon fusion process as it is colorless.

The calculation of $\kappa_g$ thus becomes completely analogous to
that of $\kappa_{\gamma}$ in the previous section, except that the $W$ boson
loop is absent.
So we obtain that the ratio of the gluon fusion production cross section
$\sigma(gg \to H)$ in our model to the usual SM gluon fusion production
cross section $\sigma_{SM}(gg \to H)$, which is equal to the square of the
scale factor $\kappa_g$, is given by
\be
\frac{\sigma(gg \to H)}{\sigma_{SM}(gg \to H)}
=\kappa_g^2 = \left(1 - \frac{K}{\frac 43 A_t}\right)^2.
 \lb{20e} \ee
Thus the expression for $\kappa_g$ in our model is just
\be
\kappa_g = 1 - \frac{K}{0.46}.
 \lb{20f} \ee

\section{Predicted correlation between $\kappa_{\gamma}$ and $\kappa_g$}

As shown in the previous two sections, the scale factors $\kappa_{\gamma}$ and
$\kappa_g$ for the Higgs diphoton decay and the Higgs gluon fusion production
amplitudes respectively are given by the expressions:
\be
\kappa_{\gamma} = 1 + \frac{K}{1.62},  \qquad
\kappa_g = 1 - \frac{K}{0.46}.
\lb{kappas} \ee
Both scale factors are expressed in terms of the same amplitude $K$,
which is just the sum of the amplitudes (\ref{35b}) for the
contributions of our
new bound states $S$ and $F$ to Higgs diphoton decay. Thus our model
predicts a correlation between the scale factors corresponding to
a line in the $(\kappa_{\gamma}, \kappa_g)$ plane parameterised by the
value of the amplitude $K$. This correlation just depends on the fact
that the constituents of our bound states are top and anti-top quarks,
which couple to photons and the analogous gluons in a quite fixed ratio
(corresponding to an electric charge of 2/3 and a color triplet coupling).

Now the amplitude $A_F$ is positive, while the amplitude $A_S$
is negative. However, as pointed out in section 2, the 11 quark
bound state $F$ is much heavier than the 12 quark bound state $S$
and, as we confirm in section 7, $A_F$ is numerically negligible
compared to $A_S$. So we conclude that
\be
K = 12 A_S > 0.
\ee
It follows that $\kappa_{\gamma}$ is positive in our model, but
$\kappa_g$ can be negative for  $K > 0.46$. Experimentally only
the magnitude of the scale factors can be determined. So, in
Fig.~3, we show our model predictions for $\kappa_{\gamma}$
and $|\kappa_g|$ with $K > 0$, as a bent line, compared to the
CMS and ATLAS results.

\bfi\centering
\includegraphics[height=100mm,keepaspectratio=true,angle=0]
{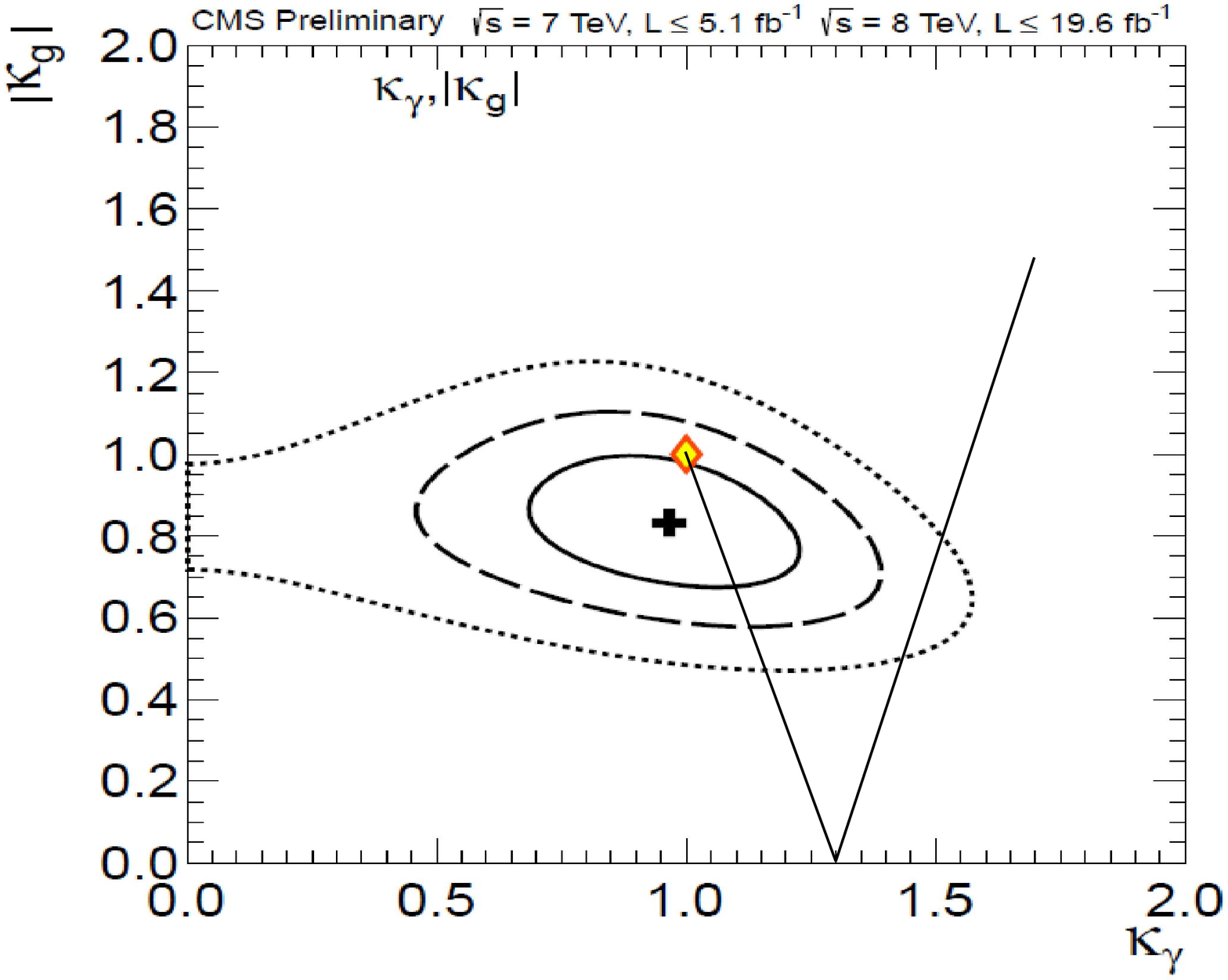}
\includegraphics[height=100mm,keepaspectratio=true,angle=0]
{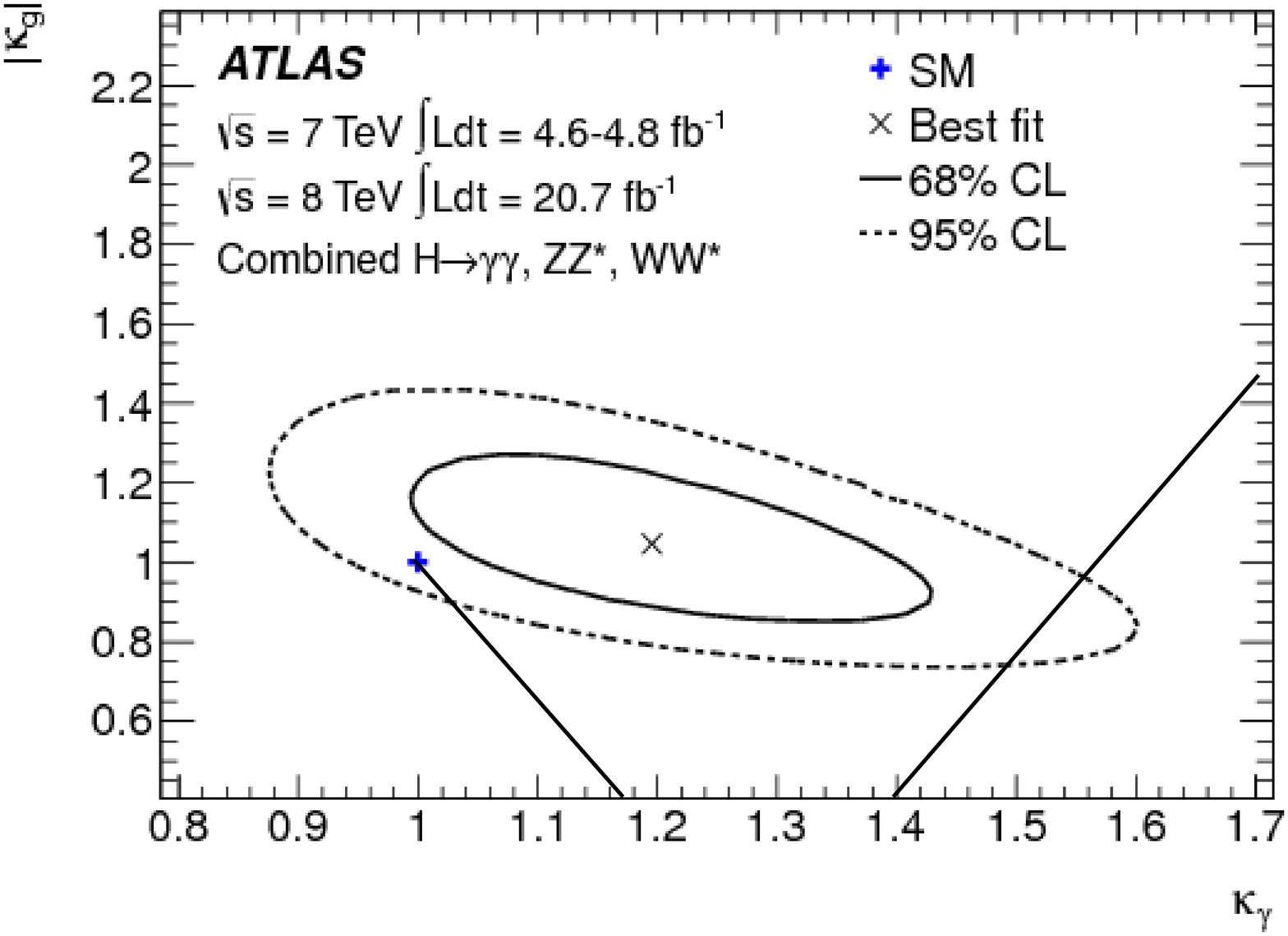}
\caption{CMS ( 68\%, 95\% and 99.7\% CL) and ATLAS (68\% and 95\% CL)
Collaboration likelihood contours for the scale factors $\kappa_{\gamma}$
and $\kappa_g$. The bent line represents the correlation between
$\kappa_{\gamma}$ and $\mid \kappa_g \mid$ predicted by our model
for positive values of the amplitude $K$.}\lb{F4}\efi

Clearly the CMS and ATLAS results are consistent with our model
at the 2 standard deviation level for small values of $K$ and
$(\kappa_{\gamma}, \kappa_g)$ close to the usual SM values
(1,1). We estimate upper limits for $K$ at the 3 standard
deviation level in this case, corresponding to the first part
of the bent line in Fig.~3 with $\kappa_g >0$,
to be $K < K_{CMS} = 0.24$ and $K < K_{ATLAS} =0.12$ for the
CMS and ATLAS data respectively\footnote{We note that $K < 0.24$
corresponds to $\kappa_{\gamma} < 1.15$ and $K < 0.12$ corresponds
to $\kappa_{\gamma} < 1.074$.}.
However our model is also consistent with the ATLAS
results at the 2 standard deviation level with values of $K \approx 0.8$
and $(\kappa_{\gamma} \approx 3/2, \kappa_g \approx -3/4)$,
while only at the 3 standard deviation level with the CMS
results.

In the following sections we estimate, as best we can,
the value of the amplitude $K$ in our model by evaluating the
loop diagrams in Fig.~2 for the bound states $S$ and $F$.

\section{Form factor corrections}

Our intuition suggests that the bound states $S$ and $F$ can in first
approximation be treated as ``fundamental" particles in loop calculations,
due to the fact that they are so tightly bound. For this reason we
introduce additional Feynman rules with propagators for the bound states
and some appropriate coupling vertices. In principle, it should not be
needed to do so, since at a very high order the long series of Feynman
diagrams should produce the effects of bound states, but in practice
we can hope to get meaningful and correct contributions using the
propagators and vertices for the bound states. However the bound
states are not really small compared to their Compton wavelengths
in the Bohr hydrogen atom like model we use in appendix B to estimate
the polarizability of $S$. Thus we expect to need form factors modifying the vertices
or the propagators to be used, when the 4-momenta become of the order of
the inverse radius of the bound state. We therefore propose
the following crude rule for taking the structure of the bound states into
account in Feynman loop integrals:
When considering a bound state $i$ with an extension or radius
$r_0 = \sqrt{<r_i^2>}$, we introduce a form factor behaving like
\be \mathfrak F \approx \exp (\frac 16 <r_i^2> q^2), \lb{36} \ee
where $q$ is the four momentum relevant for the propagator (or effective
vertex) in question. After a Wick rotation the loop four momentum  $q$
will be space-like, so that $q^2 < 0$ and the contribution of the
Feynman diagram is numerically damped.

So, in order to improve our estimate of the digrammatic contribution
of bound states running around a loop, we suppose that exponential form
factors (\ref{36}), one for each propagator, come in multiplying the
integrand of the Feynman diagram. This means that we include in the
$q$-loop integral an extra factor
\be {\mathfrak F}_{loop} = \exp(\frac 16<r^2>q^2), \lb{36a} \ee
where $<r^2>=\sum_i <r_i^2>$ is a quadratic sum of the radii of the
particles, denoted by $i$, occurring around the loop.

We estimate the radii of our bound states using the Bohr model structure
of appendix B.
So we use the value
\be
<r^2_S> = <r_F^2> = 3 a_B^2
\ee
for the spin-0 and spin-1/2 bound states $S$ and $F$, while for the spin-1
excited state $S_1^*$ we use
\be
<r^2_{S_1^*}> = 30a_B^2
\ee
for the 2p orbit.
The Bohr radius $a_B$ is given by Eq.~(\ref{A4}) of appendix B.
Hence the squared radius parameter appropriate to the fermionic
loop diagram of Fig.~2a for the bound state $F$ contribution is
$<r^2> = 3<r_F^2> = 9a_B^2$. While in the bosonic loop diagram
of Fig.~2b for the bound state $S$ contribution we have
$<r^2> = 2<r_S^2> +<r_{S_1^*}^2> = 36a_B^2$.

\section{Evaluation of bound state contributions}

In order to compare our model predictions (\ref{kappas}) for the scale factors
$\kappa_{\gamma}$ and $\kappa_g$ with the CMS and ATLAS experimental data
in Fig.~3, we need the value of the amplitude
$K = -(12A_S + \frac{44}{3}\frac {m_t}{m_F}A_F)$. So we now turn to the
evaluation of the amplitudes $A_S$ and $A_F$.

Firstly we consider the amplitude $A_S$ introduced in
Eq.~(\ref{25}) to parameterise the contribution of Fig.~2b to the
Higgs diphoton decay. The dipole moment transition vertex $SS_1^*\gamma$
is defined in Eq.~(\ref{A15}) of appendix B, where we introduce a
scalar field $\phi_S$ for the bound state $S$ and an antisymmetric
tensor field $V_{\mu\nu}$ for the spin-1 bound state $S_1^*$. The dipole
moment tansition amplitude $d_1$ is estimated, in appendix B, using a
calculation analogous to that of the dipole moment matrix elements
for the hydrogen atom in atomic physics.

The mass $m_{S_1^*}$ of the excited bound state $S_1^*$ is expected to be
much larger than the mass $m_S$ of the bound state $S$ and the Higgs mass
$m_H$. So, in evaluating the amplitude for the loop diagram of Fig.~2b in
appendix C, we actually take the large $m_{S_1^*}$ limit and integrate out the
$S_1^*$ field. In this way, we effectively contract the propagator and
the vertices involving the heavy bound state $S_1^*$ into a four-leg
polarization vertex as illustrated in Fig.~4. The effective interaction
corresponding to this four-leg polarization vertex is given in
Eq.~(\ref{A13}) of appendix B and the coupling constant is given
by the polarizability $\alpha_{pol.}$ of the bound state $S$. The
polarizability $\alpha_{pol.}$ is expressed in terms of the dipole
moment transition amplitude $d_1$ in Eq.~(\ref{A8}) of appendix B. Using
the values of these parameters motivated by our atomic physics model
of the bound states and a loop form factor
 ${\mathfrak F}_{0}(q^2) = \exp(6 a_B^2q^2)$ from section 6, we evaluate
the Feynman amplitude for Fig.~2b in appendix C. The following expression
(\ref{B13a}) is obtained:
\be
-12A_S \approx \frac{2^{21}}{3^{13}}\frac{m_t^2}{m_S^2} = 1.3\frac{m_t^2}{m_S^2}.
\lb{31a}
\ee

\bfi\centering\includegraphics[height=50mm,keepaspectratio=true,angle=0]
{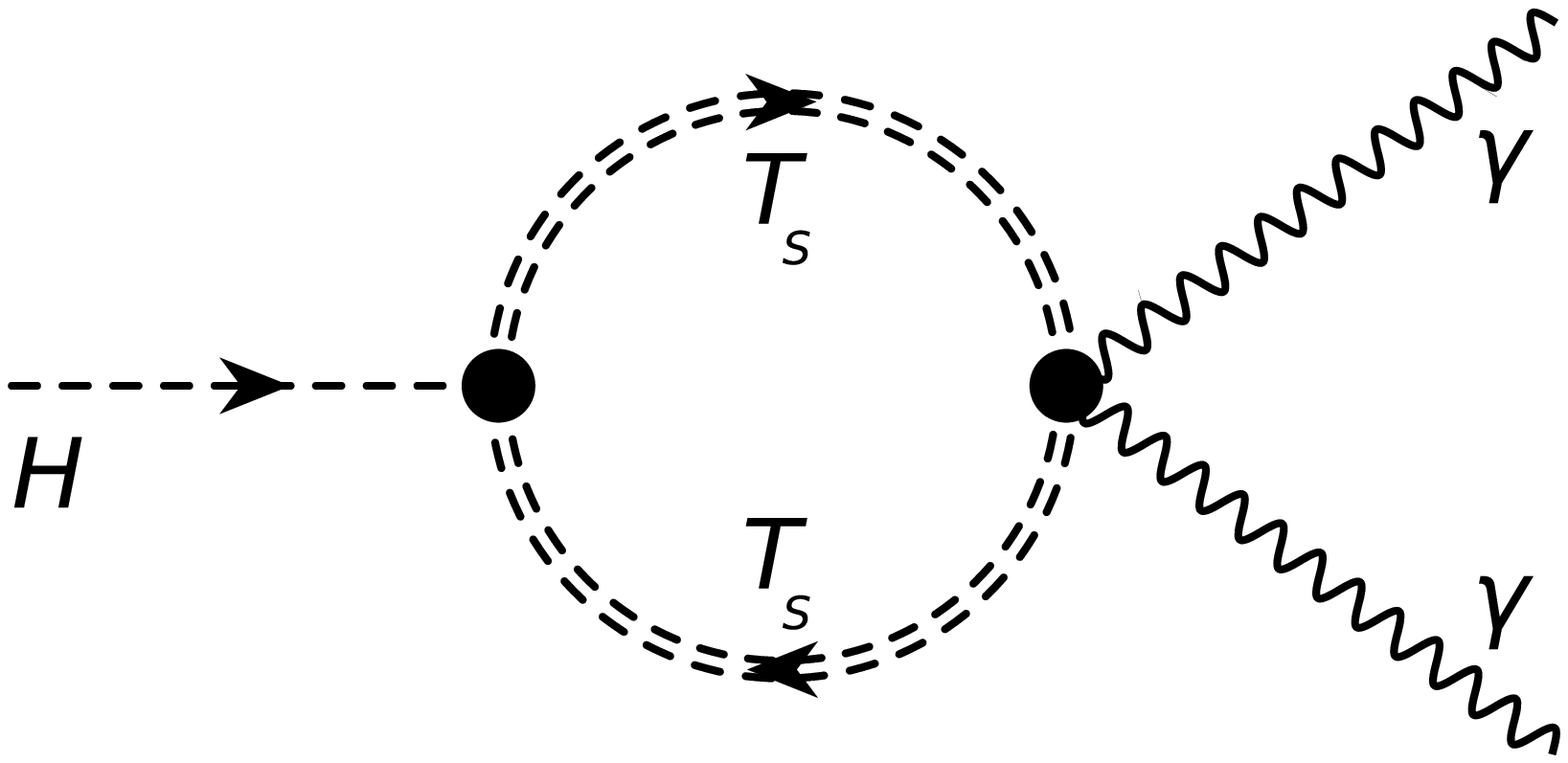}\caption{The effective one loop diagram describing the
contribution of $T_s$ -- the scalar bound state S --
to the diphoton decay of the Higgs boson.} \lb{F3}\efi

The Feynman amplitude for the loop diagram of Fig.~2a for the spin-1/2 bound
state $F$ is considered in appendix D. If $F$ were truly an elementary particle,
the value of the amplitude $A_F$ in Eq.~(\ref{25}) would be given by
$A_{1/2}(\tau_F) \approx 1/3$.
However this value is substantially reduced in the presence of the loop
form factor ${\mathfrak F}_{1/2}(q^2) = \exp(3 a_B^2q^2/2)$ from section 6
for the bound state $F$. The resulting Feynman amplitude is estimated in
appendix D, using the approximation that the mass $m_F$ of the 11 quark bound
state $F$ is much larger than the Higgs mass $m_H$. The following
expression (\ref{D4}) is obtained:
\be
\frac{44}{3} A_F  = 0.034\left (\frac{m_t}{m_F} \right )^3. \lb{Af}
\ee

The mass of the spin-1/2 bound state $F$ is expected to be
much larger than that of the scalar bound state $S$ and the top
quark mass. So we can neglect the contribution
(\ref{Af}) of the bound state $F$ to the amplitude $K$ compared
to the contribution (\ref{31a}) of the bound state $S$:
$-12A_S \gg \frac{44}{3} \frac{m_t}{m_F} A_F$.
Therefore we finally end up with the following expression for the
amplitude $K$:
\be
K = -12A_S \approx 1.3\frac{m_t^2}{m_S^2}. \lb{KAS}
\ee

\section{Results}

Our most secure result is the predicted correlation (\ref{kappas})
between the scale factors $\kappa_{\gamma}$ and $\kappa_g$. Using
our atomic physics based model for the four-leg polarization vertex
and form factor corrections, we have evaluated the loop diagram of
Fig.~4 to obtain our estimate (\ref{KAS}) for the amplitude $K$ and
hence for the scale factors:
\be
\kappa_{\gamma} = 1 + 0.80\left(\frac{m_t}{m_S}\right)^2,  \qquad
\kappa_g = 1 - 2.8\left(\frac{m_t}{m_S}\right)^2.
\lb{R1}
\ee

From the CMS and ATLAS data in Fig.~3, we estimated in section 5
the corresponding 3 standard deviation upper limits on the amplitude $K$
(for positive values of $\kappa_g$) to be $K_{CMS} = 0.24$ and
$K_{ATLAS} = 0.12$ respectively. These limits lead, using
Eq.~(\ref{KAS}), to the following estimates of the 3 standard deviation
lower limits on the mass of our bound state $S$:
\be
\mbox{CMS}\ 3\sigma \ \mbox{limit}: \quad  m_S > 400 \ \mbox{GeV},
\qquad
\mbox{ATLAS}\ 3\sigma \ \mbox{limit}: \quad  m_S > 570 \ \mbox{GeV}.
\lb{R2}
\ee

There is also a region of parameter space with negative values of
$\kappa_g$ and $K \approx 0.8$, lying on the second part of the bent
line in Fig.~3, which is consistent with the CMS and
ATLAS data at better than 3 standard deviations. In fact the corresponding values
of the scale factors
\be
 \kappa_{\gamma} \approx 1.5, \qquad \qquad \kappa_g \approx -0.75
 \lb{R3}
\ee
are consistent with the CMS data at the $2.5\sigma$ level and
with the ATLAS data at the $1.5\sigma$ level. We note that the latter
fit to the ATLAS data is as good as the usual SM fit
($\kappa_{\gamma}$ = 1, $\kappa_g$ =1). From Eq.~(\ref{KAS}), we
estimate the mass of the bound state $S$ giving rise to the
scale factors (\ref{R3}) in our model to be:
\be
m_S \approx 220 \ \mbox{GeV}.
\lb{R4}
\ee

\section{Conclusion}

In this paper we have investigated further the proposed existence of a
$6t + 6\bar t$ scalar bound state $S$ and a $6t + 5\bar t$ spin-1/2
bound state $F$, which are so strongly bound and exceptionally light
that they effectively act as elementary particles to first approximation.
We have estimated the loop corrections arising therefrom and the consequent
scaling of the usual SM amplitudes for Higgs diphoton decay and for Higgs
production via gluon fusion. Our predictions are rather uncertain due to
the approximations we make and, in particular, to the large uncertainty
on the mass $m_S$ of the bound state $S$.

We have compared our results for the scale factors $\kappa_{\gamma}$ and
$\kappa_g$ with the LHC data of Fig.~3. We find two regions of
allowable parameter space. They correspond to masses $m_S > 400$ GeV
for CMS or $m_S > 570$ GeV for ATLAS data and $m_S \sim 220$ GeV
respectively.

Our most robust prediction follows from the constituents of the bound
states $S$ and $F$ being just top and anti-top quarks. Consequently there is
a linear relationship between $\kappa_{\gamma}$ and $\kappa_g$,
resulting from the effective proportionality between the relevant
diagrams of Fig.~2 coupling the Higgs to two photons and the
corresponding diagrams coupling it to two gluons. This relationship is
represented by the bent line in Fig.~3 and is parameterised by
the amplitude $K$ in Eq.~\ref{kappas}. The diphoton decay width is
expressed in terms of the amplitude $K$ and the usual SM
contributions from $W^{\pm}$ and top quark loops as follows:
\be
 \Gamma(H\to\gamma\gamma) =
 \frac{\alpha^2m_H^3}{16\pi^3v^2} |A_W(\tau_W) + \frac 43 A_t(\tau_t)
  - K|^2.
    \lb{1con} \ee
An analogous expression for the gluon fusion production cross section
$\sigma(gg \to H)$ leads to Eq.~\ref{20e}.

The dominant contribution to the amplitude $K$ is provided by the loop diagram
of Fig.~2b involving the color singlet self-conjugate 1s bound state $S$
of 6 top and 6 anti-top quarks. We treat its interaction with photons, or
quite analogously with gluons, as due to the polarizability of the state $S$
by its being virtually excited up to a spin-1 state $S_1^*$ in which one of
the quarks is excited to a 2p level. We estimate the polarizability
$\alpha_{pol.}$ using a non-relativistic Bohr model and then translate it into
a relativistic interaction using the procedure described in appendix A. The
structure of the bound states is taken into account by form factors determined
by their Bohr radii. The contribution of the spin-1/2 bound state $F$ turns
out to give a relatively small contribution to $K$ and can be neglected.

The contribution of the loop diagram of Fig.~2b to the amplitude $K$ is estimated
to be
\be
K = \frac{9 \alpha_{pol.} m_t m_S^2}{\pi \alpha}
\int_0^{\infty} \mathfrak{F}_0(q_E^2)\frac{q_E^2dq_E^2}{(q_E^2 + m_S^2)^2}
\approx 1.3\frac{m_t^2}{m_S^2},
\lb{2con}
\ee
where $\mathfrak{F}_0(q_E^2)$ is the form factor, discussed in section 6,
for the $S$ bound state contribution. The bound state $S$ is supposed to be
much lighter than $12 m_t \sim 2$ TeV, but its mass $m_S$ is otherwise very
uncertain. Realistically we would say our model requires $m_S < 6 m_t$ and
hence $K > 0.036$. This means that the deviations from unity of the scale
factors $\kappa_{\gamma}$ and $\kappa_g$ should be greater than $2.2 \%$
and $7.9 \%$ respectively; otherwise our model is falsified. In a recent
paper \ct{52} we made a crude estimate of $m_S \approx 260$ GeV for the
mass, which  happens to be close to the mass of 220 GeV corresponding to
the allowed region of parameter space with negative $\kappa_g$.

We emphasize again that this paper concerns the {\em pure} Standard Model.
Usually one thinks that, apart from the strongly interacting gluons, the
Feynman diagram expansion converges rather quickly. However if there are
bound states, then in certain small regions it could be that the diagrams
add up and produce a pole where the bound state is close to being on-shell.
It is the existence of such a bound state that we propose here and, strictly
speaking, are making a speculative correction to the usual SM calculation
rather than introducing new physics. We do, however, suggest that the
SM coupling constants are fine-tuned, by the so-called Multiple Point
Principle, so as to produce a number of vacua with the same energy
density. For this principle of degenerate vacua to be realized a bound state
$S$ of 6 top and 6 anti-top quarks is needed to be exceptionally light.
An analysis \ct{10} of the complicated dynamics involved in the strong
binding of the 12 quark state supports this conclusion, but the result
has been disputed \ct{15,16,17,18} and more accurate but difficult
calculations are needed.

Also it would be desirable to improve our Bohr model calculation of
the polarizability of the bound state $S$ and of the form factors, by
introducing the extra binding effects due to the exchanges of gluons,
$W$'s etc and the other corrections discussed in Ref.~\ct{10}.
It must though be admitted that this would mainly help fixing the
value of the radius of the bound state, which we have seen may only
influence our predictions mildly.

\section{Acknowledgments}

H.B.~Nielsen thanks the Niels Bohr Institute for emeritus status
and the University of Helsinki for a visit.
L.V.~Laperashvili deeply thanks the Niels Bohr Institute
(Copenhagen, Denmark) for hospitality and financial support, and
Roman Nevzorov (ITEP, Moscow) for very useful discussions.
C.R.~Das greatly thanks Prof. Utpal Sarkar
for financial support through
J.C. Bose fellowship.
C.D.~Froggatt thanks Glasgow University and the Niels Bohr Institute
for hospitality and support.

\section*{Appendix A. Non-relativistic formulation of vertices}

In the present article we introduce a standard procedure to
translate from a non-relativistic expression for vertices involving
our $6t + 6\bar t$ spin-0 and spin-1 bound states $S$ and $S_1^*$
into a relativistic expression. We make a relativistic ansatz for
a vertex and adjust the coefficient so that it matches the
non-relativistic one. The basic rule followed is
to use the usual relativistic normalization for, say, a scalar
field $\int (2E)\phi^{\dagger}\phi d^3x =1$, and make the
approximation $2E \approx 2m$. For example when we consider the
$HSS$ vertex, treating the two scalar bound states $S$ as
non-relativistic resting objects while $H$ is the interacting
Higgs field, we get an extra factor
$(\sqrt{2m_S})^2$ into the expression for the relativistic
coupling constant (\ref{34}):

\begin{equation}
G_{HSS} = 12\frac{g_t}{\sqrt{2}} \cdot 2m_S.
\end{equation}

In appendix B we transform the non-relativistic
polarization Hamiltonian (\ref{A1}) into a relativistic expression
(\ref{A13}) for the four-leg polarization vertex appearing in Fig.~4.
We introduce the square of a scalar quantum field $\phi_S$
describing the scalar bound state $S$ into the Hamiltonian density (\ref{A12})
together with just two factors of $\sqrt{2m_S}$ to provide the
correct relativistic normalization. We also introduce a normalization
factor $\sqrt{2m_{S_1^*}}$ for the antisymmetric tensor field
$V_{\mu\nu}$ associated with the spin-1 bound state $S_1^*$.

\section*{Appendix B: On describing the polarizability}

In this appendix we discuss the interaction of
$\gamma$-quanta with the bound states $S$ and $S_1^*$
given by the diagram of Fig.~2b and the effective interaction
obtained by integrating out $S_1^*$ and given by the diagram of
Fig.~4. This effective interaction, estimated in the
non-relativistic limit of the bound state $S$ being
essentially at rest, is described by the following
polarization term in the Hamiltonian:
\be H_{pol.}= - \frac{1}{2}\alpha_{pol.} \vec{E}^2, \lb{A1} \ee
where the constant $\alpha_{pol.}$ is called the polarizability. A
model of polarizability is given in Refs.~\ct{32,33}.

When a neutral bound state is placed in an electric field
$\vec{E}$, the positively charged part of it gets pulled in the direction
of the field, and the negative cloud gets pulled in the direction
opposite to the field. This results in the centres of positive and
negative charge no longer being in the same place, so the bound
state obtains a small dipole moment $\vec d$. Experimentally, it
is found that for small fields this dipole moment is
approximately proportional to the applied field:
\be  \vec d = \alpha_{pol.}\vec{E}. \lb{A2} \ee

Let us now consider the dipole moment transition amplitude $d_1=qR_1$
for the bound state $S$, where $q=+2/3|e|$ and $R_1$
is the radius of transition  from $S$ to the $S^*_1$-particle,
which is a $2p$ excited state of the spin-0 $6t+6\bar t$ bound state $S$.
We make an estimate of the value of $d_1$ based on the atomic physics
calculation of the dipole moment matrix elements of the hydrogen atom.
%
%
%
%
%
According to the atomic physics terminology, we have:
\be R_1= <2,1,0|z|1,0,0>= \sqrt 2 \frac{2^7}{3^5}a_B\approx
0.744a_B, \lb{A3} \ee
where $a_B$ is the radius of the Bohr Hydrogen-atom-like bound
state $S$, in which we approximate the interaction of each top quark
by a central $1/r$ potential (neglecting the Higgs mass) while only
half of the other 11 particles\footnote{We note that our
estimate of the loop amplitude $A_S$ in appendix C is insensitive
to the number $Z$ of quarks we take to be in the ``nucleus"
and therefore to the radius $a_B$.}
are present in the ``nucleus" \cite{11}. The variable $z$ denotes the
position coordinate in an axial coordinate system.
Therefore we have:
\be a_B= \frac{4\pi}{\frac{11}{2}(g_t^2/2)m_t} \approx
\frac{5.2}{m_t}. \lb{A4} \ee
for a running top quark Yukawa coupling $g_t \approx 0.935$.

In the philosophy that the excited state $S_1^*$ of the ground
state $S$ is a 2p state with an excitation energy equal to 3/4
of the ground state binding energy $E_{Ry}$, we obtain \ct{33}
the contribution of each of the 12 quark constituents to the
polarizability of $S$ to be:
\be \alpha_{pol.}= \frac{2d_1^2}{\Delta E}
= 2q^2\frac{|<1s|z|2p>|^2}{ \Delta E}
                              \lb{A8} \ee
with
\be \Delta E = \frac 34 E_{Ry}.   \lb{A9} \ee
Here we used Heaviside-Lorentz units instead of the Gaussian units
of Ref.~\ct{33}. In our case we have:
%
%
\be  E_{Ry} = \frac{11g_t^2}{32\pi a_B}. \lb{A10} \ee
The value of the transition amplitude $<1s|z|2p>$ is given by
(\ref{A3}), and Eq.~(\ref{A8})
leads to the following result for the polarizability:
%
%
\be
\alpha_{pol.}=4\pi\frac{e^2}{11g_t^2}\frac{2^{23}}{3^{13}}a_B^3.
                           \lb{A11} \ee

The polarization Hamiltonian (\ref{A1}) is not relativistically
invariantly as written and one should also note that there is no analogous
magnetic interaction in the rest system of the bound state.
So firstly we rewrite (\ref{A1}) as a Hamiltonian density in a quantum
field theory like form, with a relativistically normalized self-conjugate
scalar field $\phi_S$ describing the scalar bound state $S$
as explained in appendix A:
\be {\cal H}_{polarization}=- \frac{\alpha_{pol.}}{2}
(2m_S)|\phi_S|^2 \frac{1}{2} \vec{E}^2. \lb{A12} \ee
Then we perform an averaging over all possible Lorentz
transformations of a corresponding Lagrangian density term, so as
to obtain a Lorentz invariant expression by the replacement:
\be
    \frac 12 {\vec E}^2 \to \frac 12 ({\vec E}^2 - {\vec B}^2) \to
    \frac 14 F_{\mu\nu}(x) F^{\mu\nu}(x), \lb{A14} \ee
where
\be
\vec{E}|_i = F^{0i},\qquad
\vec{B}|_i = \frac 12\epsilon_{ijk} F^{jk}.
\ee
Thus finally we obtain a covariant expression for the polarization
Lagrangian density:
\be {\cal L}_{pol.}(x) = m_S\frac{\alpha_{pol.}}{4} F_{\mu\nu}(x)
F^{\mu\nu}(x)|\phi_S(x)|^2.  \lb{A13} \ee
which replaces the polarization Hamiltonian as
given by (\ref{A1}) for a single bound state at rest and describes the
four-leg polarization vertex in Fig.~4.

The vertex $SS^*_1\gamma$ in the diagram of Fig.~2b is described
by the interaction of the type
\be
       \frac12d_1\sqrt{(2m_S)(2m_{S_1^*}})
\phi_SF^{\gamma}_{\mu\nu}V^{\mu\nu}, \lb{A15}   \ee
where $V_{\mu\nu}= - V_{\nu\mu}$ is an antisymmetric tensor field
describing the vector particle $S^*_1$ and we have introduced the
normalization factor $\sqrt{(2m_S)(2m_{S_1^*}})$.
In Eq.~(\ref{A15}), an extra factor of 1/2 compensates for the
fact that there is a double counting by the
summation over the anti-symmetrized indices (here $\mu,\nu$). The
effective field, called $V_{\mu\nu}$, which describes the excited
bound state $S_1^*$, is only thought to be very crudely treated,
and we thus only write here the main kinetic and mass
terms for this field:
\be {\cal L}(x) = \frac{1}{4}
\partial_{\rho} V_{\mu\nu}
\partial^{\rho}V^{\mu\nu} - \frac{1}{4}m_{S_1^*}^2V_{\mu\nu}V^{\mu\nu}-
 d_1\sqrt{m_Sm_{S_1^*}}
\phi_SF^{\gamma}_{\mu\nu}V^{\mu\nu}. \lb{A16} \ee
We do not go into detail with the terms
destined to leave only the components of polarization properly
oriented with respect to the four momentum of the $S_1^*$-particle
 -- terms which contribute to the terms proportional to the momentum in the
propagator (see (\ref{B3}) and (\ref{B4}) below), which we shall
ignore in Appendix C.

\section*{Appendix C: Evaluation of the loop amplitude $\bf A_S$}

The diphoton decay rate for the Higgs is given, in terms of the
conventionally normalized Feynman integrals $I_{W,t,F,S}$
corresponding to the Feynman diagrams of Fig.~2, by
\be \Gamma(H\to \gamma\gamma)= \frac {1}{32\pi m_H}|I_W + I_t + I_F
+I_S|^2.       \lb{B1} \ee
%
%
A symmetry factor of 1/2 has been included in the phase space
for the two identical photons and there is an implicit sum over
their polarizations.

The usual SM expressions for $I_W$ and $I_t$ can be extracted
from Section 3. We estimate $I_F$ in appendix D and find that
it can be ignored compared to the other contributions $I_{W,t,S}$.

The last term from (\ref{B1}), which comes from the diagram
of Fig.~2b, is more cumbersome. Indeed we shall take the limit
of a very heavy spin-1 bound state $S_1^*$ and effectively
integrate it out. We thereby replace the part of the diagram
around the $S_1^*$-propagator by a single four-vertex, representing
the polarizability of the bound state $S$ by virtual excitation
to the $S_1^*$ state. Then we effectively have a diagram with
just two vertices that appears when the $S_1^*$-propagator and
the vertices attached at each end are contracted into one
vertex with two incoming scalars $S$ and two outgoing photons.
The corresponding diagram is given in Fig.~4.

Let us now estimate the Feynman integral $I_S$ for the loop diagram
of Fig.~2b, using the vertex (\ref{A15}) from appendix B and the
coupling $G_{HSS}$ of the Higgs field to the self-conjugate scalar
bound state $S$.
Taking into account that we have 12 different states of the
top quark in the bound state $S^*_1$ (top quark and anti-top quark
with 3 colors and 2 spin states), each of which can be excited
into the 2p level, we obtain the following
integral:
$$ I_S = -i12G_{HSS}d_1^2m_S m_{S_1^*} \int \frac{d^4q}
{{(2\pi)}^4}{\mathfrak F}_0(q^2)\frac {1}{(q+k_1)^2- m_S^2}\cdot
\frac {1}{(q-k_2)^2- m_S^2}\times $$ \be \frac
{\Pi_{\mu\nu\mu'\nu'}}{q^2 - m^2_{S^*_1}}\times [k_{1\mu}
\epsilon_{\nu}(k_1) - k_{1\nu} \epsilon_{\mu}(k_1)][k_{2\mu'}
\epsilon_{\nu'}(k_2) - k_{2\nu'} \epsilon_{\mu'}(k_2)]. \lb{B2}
\ee
Here
\be \Pi_{\mu\nu\mu'\nu'} = g_{\mu\mu'} g_{\nu\nu'}  -
\frac{q_{\mu}q_{\mu'}}{ m_{S^*_1}^2} g_{\nu\nu'} -
\frac{q_{\nu}q_{\nu'}}{ m_{S^*_1}^2} g_{\mu\mu'} +
\frac{q_{\mu}q_{\mu'}q_{\nu}q_{\nu'}}{ m_{S^*_1}^4}. \lb{B3} \ee
and ${\mathfrak F}_0(q^2)$ is the form factor (\ref{36a})
for the bound states propagating round the loop as
estimated in section 6:
\be {\mathfrak F}_0(q^2) = \exp(\frac 16<r^2>q^2),  \ee
where
\be
<r^2> = 2<r_S^2> + <r_{S_1^*}^2> = 36a_B^2. \lb{B3a}
\ee

We now assume the bound state $S_1^*$ is very heavy:
$m_{S^*_1}\gg m_S,\,m_H/2$, so we can replace the
propagator of the bound state $S_1^*$ using the approximation:
\be
 \frac {\Pi_{\mu\nu\mu'\nu'}}{q^2 - m^2_{S^*_1}}\approx
 - \frac{g_{\mu\mu'}g_{\nu\nu'}}{m^2_{S^*_1}}.  \lb{B4}
\ee
If we further make the identification
\be \Delta E = m_{S_1^*} - m_S \approx m_{S_1^*}
\ee
in the expression (\ref{A8}) for the polarizability, we obtain
the following result:
\be d_1^2 = \frac 12 \alpha_{pol.}m_{S^*_1}.   \lb{B4a} \ee
Then, using the approximation (\ref{B4}) and the relations
(\ref{B4a}) and (\ref{34}), we obtain:
$$ I_S
=  i\frac{144\alpha_{pol.}}{v}m_tm_S^2\int
\frac{d^4q}{{(2\pi)}^4}{\mathfrak F}_0(q^2)\frac {1}{(q+k_1)^2-
m_S^2}\cdot \frac {1}{(q-k_2)^2- m_S^2}\times $$ \be  [(2k_1\cdot
k_2)(\epsilon_1\cdot \epsilon_2) - 2(k_1\cdot \epsilon_2)(k_2\cdot
\epsilon_1)]. \lb{B5} \ee
We note that we get the same expression for $I_S$ from Fig.~4,
using the effective interaction (\ref{A13}) supplemented with
the form factor ${\mathfrak F}_0(q^2)$. This provides a useful
consistency check on our calculation and also explains why the
mass $m_{S_1^*}$ cancels out from our expression (\ref{B5})
for $I_S$.

We now make the approximation of neglecting the photon momenta
$k_{1,2}$ as small compared to $m_S$ in the propagators of the
expression (\ref{B5}) and obtain:
\be I_S = i\frac{144\alpha_{pol.}}{v}m_tm_S^2\int
\frac{d^4q}{{(2\pi)}^4}{\mathfrak F}_0(q^2)\frac {1}{(q^2-
m_S^2)^2} [(2k_1\cdot k_2)(\epsilon_1\cdot \epsilon_2) - 2(k_1\cdot
\epsilon_2)(k_2\cdot \epsilon_1)].  \lb{B6} \ee
After the Wick rotation $q^0 = iq_E^4$, $q^2 = -q_E^2$ we have:
\be I_S = - \frac{144\alpha_{pol.}}{v}m_tm_S^2\int
\frac{d^4q_E}{{(2\pi)}^4}{\mathfrak F}_0(q_E^2)\frac {1}{(q_E^2 +
m_S^2)^2}[(2k_1\cdot k_2)(\epsilon_1\cdot \epsilon_2) - 2(k_1\cdot
\epsilon_2)(k_2\cdot \epsilon_1)].  \lb{B7} \ee
The symbol $E$ occurring as an index in (\ref{B7}) means that
we are now using a Euclidean metric.

Taking into account that $$\frac{d^4q_E}{{(2\pi)}^4}=\frac
{q_E^2dq_E^2}{16\pi^2},$$ we obtain:
\be I_S = - \frac{144\alpha_{pol.}}{v}m_tm_S^2\int
\frac{q_E^2dq_E^2}{16\pi^2}{\mathfrak F}_0(q_E^2)\frac {1}{(q_E^2
+ m_S^2)^2}[(2k_1\cdot k_2)(\epsilon_1\cdot \epsilon_2) -
2(k_1\cdot \epsilon_2)(k_2\cdot \epsilon_1)].  \lb{B7a} \ee

We can also choose the polarizations $\epsilon_i$ for the photons
to be orthogonal to both photon momenta $k_1$ and $k_2$.
From the kinematics it is easy to see that
$2k_1\cdot k_2 = m_H^2$ on shell, and so we have:
\be I_S =  - \frac{m_H^2}{16\pi^2}(\epsilon_1\cdot \epsilon_2)
\frac{144\alpha_{pol.}}{v}m_tm_S^2\int_0^{\infty}{\mathfrak
F}_0(q_E^2) \frac {q_E^2dq_E^2}{(q_E^2 + m_S^2)^2}. \lb{B8} \ee
Using the notation:
\be A_0 = \int \frac{{\mathfrak F}_0(q_E^2)q_E^2dq_E^2}{(q_E^2 +
m_S^2)^2} = \int_0^{\infty}\frac{{\mathfrak F}_0(y)ydy}{(y+1)^2} =
\int_0^{\infty} \frac{e^{-y/y_0}ydy}{(y+1)^2}, \lb{B9} \ee
where
\be y=\frac{q_E^2}{m_S^2},\quad  y_0 = \frac{6}{m_S^2<r^2>},\quad
{\rm{and}}\quad {\mathfrak F}_0(y)=e^{-y/y_0}, \lb{B10} \ee
we have:
\be I_S =  - \frac{m_H^2}{16\pi^2}(\epsilon_1\cdot \epsilon_2)
\frac{144\alpha_{pol.}}{v}m_tm_S^2A_0. \lb{B8a} \ee
%
%
%

Now, by comparison of Eqs.~(\ref{35}) and (\ref{B1}), we have
\be
I_S = \frac{\sqrt{2}\alpha m_H^2}{\pi v}12A_S. \lb{B12a}
\ee
Also, when the expression (\ref{B8a}) for $I_S$ is inserted into
(\ref{B1}), summing over the four polarization possibilities
for the diphoton final state gives
$|\epsilon_1\cdot \epsilon_2|^2 =2$. So we can rewrite (\ref{B12a})
in the form
\be
I_S = \frac{\alpha m_H^2}{\pi v}\epsilon_1\cdot \epsilon_2 12A_S.
\lb{B12b}
\ee
It then follows from (\ref{B8a}) that
\be 12A_S = - \frac{36\alpha_{pol.}}{4\pi \alpha} m_tm_S^2 A_0.
\lb{B13} \ee

The radius parameter appearing in the form factor
${\mathfrak F}_0(q^2)$ is estimated in section 6 and
given in Eq.~(\ref{B3a}) as $<r^2>=36a_B^2$.
Since this radius $<r^2>$ is large, it follows that $y_0$
is small:
\be
y_0 = \frac{6}{m_S^2<r^2>} = \frac{1}{6a_B^2m_S^2}
\approx\frac{1}{160}\left(\frac{m_t}{m_S}\right)^2,
\ee
and we can use the following approximation for the
integral (\ref{B9}):
\be A_0\approx  y_0^2
= \frac{1}{36a_B^4m_S^4}.
                      \lb{B14} \ee

Expressions for the Bohr radius $a_B$ and the polarizability
$\alpha_{pol.}$ of the scalar bound state $S$ are given in
terms of the top quark mass in (\ref{A4}) and (\ref{A11}) of
appendix B. Substitution of these expressions into (\ref{B13})
and (\ref{B14}) gives:
\be
-12A_S \approx \frac{4\pi}{11g_t^2}\frac{2^{23}}{3^{13}}\frac{m_t}{a_Bm_S^2}
= \frac{2^{21}}{3^{13}}\frac{m_t^2}{m_S^2} = 1.3\frac{m_t^2}{m_S^2}.
\lb{B13a}
\ee

\section*{Appendix D: Evaluation of the loop amplitude $\bf A_F$}

In this appendix we estimate the Feynman integral $I_F$,
appearing in (\ref{B1}), for the loop diagram
of Fig.~2a for the spin-1/2 bound state F, using the coupling $G_{HF\bar F}$
of the Higgs field to $F$. Including a factor of 2 to incorporate
the crossed diagram, we have:
\be
I_F = C \int \frac{d^4q}
{{(2\pi)}^4}{\mathfrak F}_{1/2}(q^2) Tr \left [
\frac{\slashed{q} + \slashed{k_1} +m_F}{(q + k_1)^2 -m_F^2}
\gamma^{\mu} \frac{\slashed{q} + m_F}{q^2 -m_F^2} \gamma^{\nu}
\frac{\slashed{q} - \slashed{k_2} +m_F}{(q + k_2)^2 -m_F^2}
\right] \epsilon_{\mu}(k_1)\epsilon_{\nu}(k_2),
\ee
where the coefficient $C$ is
\be
C = i2G_{HF\bar F}N_c Q_t^2 e^2.
\ee
Here ${\mathfrak F}_{1/2}(q^2)$ is the form factor (\ref{36a})
for the bound state $F$ propagating round the loop as
estimated in section 6:
\be {\mathfrak F}_{1/2}(q^2) = \exp(\frac 16<r^2>q^2),  \ee
where
\be
<r^2> = 3<r_F^2>  = 9a_B^2. \lb{B3b}
\ee

We now assume the bound state $F$ is very heavy:
$m_{F}\gg \,m_H/2$, so we can approximate the
propagator of the bound state $F$ by $-1/m_F$.
For example
\be
\frac{\slashed{q} + \slashed{k_1} +m_F}{(q + k_1)^2 -m_F^2}
\approx -\frac{1}{m_F}.
\ee
Then using (\ref{33}) we obtain
\begin{eqnarray}
I_F &\approx& i\frac{22m_t}{v}N_c Q_t^2 e^2 \int \frac{d^4q}
{{(2\pi)}^4}{\mathfrak F}_{1/2}(q^2)\left (-\frac{1}{m_F}
\right )^3 Tr \left[ \gamma^{\mu} \gamma^{\nu} \right ]
\epsilon_{\mu}(k_1)\epsilon_{\nu}(k_2) \\
& =& -i\frac{88m_t N_c Q_t^2 e^2}{v m_F^3}
\epsilon_1\cdot \epsilon_2\int \frac{d^4q}
{{(2\pi)}^4}{\mathfrak F}_{1/2}(q^2). \lb{D1}
\end{eqnarray}
After the Wick rotation $q^0 = iq_E^4$, $q^2 = -q_E^2$ we have:
\begin{eqnarray}
I_F &\approx& \frac{88m_t N_c Q_t^2 e^2}{v m_F^3}
\epsilon_1\cdot \epsilon_2 \int \frac{q_E^2dq_E^2}
{16 \pi^2}{\mathfrak F}_{1/2}(q^2) \\
&=& \frac{88 \alpha m_t}{3\pi v m_F^3} \epsilon_1\cdot \epsilon_2
\int q_E^2\exp (-3a_B^2q_E^2/2)dq_E^2 \\
&=& \frac{352 \alpha m_t}{27 \pi v m_F^3 a_B^4}
\epsilon_1\cdot \epsilon_2. \lb{D2}
\end{eqnarray}

Now, analogously to (\ref{B12b}), we derive the relationship between
$I_F$ and $A_F$ using Eqs.~(\ref{35}) and (\ref{B1}):
\be
I_F = \frac{\alpha m_H^2}{\pi v}\epsilon_1\cdot \epsilon_2
\frac{44}{3} \frac{m_t}{m_F} A_F. \lb{D3}
\ee
It then follows from (\ref{D2}) that
\be
\frac{44}{3} \frac{m_t}{m_F} A_F \approx \frac{352 m_t}{27m_H^2m_F^3 a_B^4}
= 0.018 \left (\frac{m_t}{m_H}\right )^2 \left (\frac{m_t}{m_F} \right )^3
= 0.034\left (\frac{m_t}{m_F} \right )^3.
\lb{D4}
\ee
where we have used the expression (\ref{A4}) for $a_B$ from appendix B.

\end{document}